\documentclass[conference]{IEEEtran}
\IEEEoverridecommandlockouts
\usepackage{cite}
\usepackage[english]{babel}

\usepackage[T2A,LGR,T1]{fontenc}
\usepackage[latin9]{inputenc}
\usepackage{array}
\usepackage{float}
\usepackage{times}
\usepackage[letterpaper, left=0.62in, right=0.62in, top=0.74in, bottom=1.06in]{geometry}

\usepackage{textcomp}
\usepackage{multirow}
\usepackage{stackrel}
\usepackage{graphicx}
\usepackage{booktabs}
\usepackage{setspace}
\usepackage[export]{adjustbox}
\usepackage{algorithm}
\usepackage{algpseudocode}
\usepackage{enumerate}
\usepackage{xcolor}
\usepackage{graphicx}
\usepackage{amssymb}
\usepackage{color}
\usepackage{fancyhdr}

\usepackage{amsmath,amssymb,amsfonts}
\def\BibTeX{{\rm B\kern-.05em{\sc i\kern-.025em b}\kern-.08em
    T\kern-.1667em\lower.7ex\hbox{E}\kern-.125emX}}

\begin{document}

\pagestyle{fancy}
\fancyhead{} 
\fancyhead[C]{\small This work has been accepted as a poster at the 2025 EuCNC \& 6G Summit}
\title{FedORA: Resource Allocation for Federated Learning in ORAN using Radio Intelligent Controllers %
    \thanks{*This work has been accepted as a poster in the 2025 EuCNC \& 6G Summit.}} 
\author{
  Abdelaziz Salama\textsuperscript{$\star$} \quad Mohammed M. H. Qazzaz\textsuperscript{$\star$} \quad Syed Danial Ali Shah\textsuperscript{$\star$} \quad Maryam Hafeez\textsuperscript{$\star$} \quad Syed Ali Zaidi\textsuperscript{$\star$}\\ 
  \textsuperscript{$\star$}School of Electrical and Electronic Engineering, University of Leeds, Leeds, UK\\
  \textsuperscript{$\star$}Corresponding author: Abdelaziz Salama (A.M.Salama@Leeds.ac.uk)
}


\maketitle

\begin{abstract}
This work proposes an integrated approach for optimising Federated Learning (FL) communication in dynamic and heterogeneous network environments. Leveraging the modular flexibility of the Open Radio Access Network (ORAN) architecture and multiple Radio Access Technologies (RATs), we aim to enhance data transmission efficiency and mitigate client-server communication constraints within the FL framework. Our system employs a two-stage optimisation strategy using ORAN's rApps and xApps. In the first stage, Reinforcement Learning (RL) based rApp is used to dynamically select each user's optimal Radio Access Technology (RAT), balancing energy efficiency with network performance. In the second stage, a model-based xApp facilitates near-real-time resource allocation optimisation through predefined policies to achieve optimal network performance. The dynamic RAT selection and resource allocation capabilities enabled by ORAN and multi-RAT contribute to robust communication resilience in dynamic network environments. Our approach demonstrates competitive performance with low power consumption compared to other state-of-the-art models, showcasing its potential for real-time applications demanding both accuracy and efficiency. This robust and comprehensive framework, enabling clients to utilise available resources effectively, highlights the potential for scalable, collaborative learning applications prioritising energy efficiency and network performance.

\end{abstract}

\begin{IEEEkeywords}
Federated Learning, ORAN, RIC, Multi-RAT, Dynamic Networks, Resource Allocation.
\end{IEEEkeywords}

\section{Introduction}

Several prior studies have focused on implementing Federated Learning (FL) \cite{mcmahan2017communicationefficient} mechanisms to support various network applications and ensure the required Quality of Service (QoS) \cite{li2023survey}; however, optimising FL performance under communication constraints and network limitations to enable efficient resource management for FL users remains an open challenge, particularly in dynamic and heterogeneous network environments.

To address these limitations, we propose FedORA, a novel FL framework that dynamically optimises communication efficiency and energy consumption by leveraging the Open Radio Access Network (ORAN) \cite{qiao2024ai}. ORAN's modular and open architecture enables intelligent control over Radio Access Technologies (RATs), allowing adaptive resource allocation. Unlike conventional FL implementations that rely on a single, often congested RAT, FedORA dynamically selects the most efficient RAT for each FL client based on real-time network conditions. This approach reduces communication bottlenecks, enhances FL reliability, and minimises power consumption.

Our key contributions are as follows:

\begin{itemize}
    \item \textbf{Dynamic RAT Selection}: We introduce an adaptive mechanism for selecting the optimal RAT for FL clients, ensuring efficient communication while minimising latency and energy consumption.
    \item \textbf{Energy-Aware Resource Management}: We develop xApps within ORAN to intelligently allocate resources, reducing unnecessary power consumption without compromising FL performance.
    \item \textbf{Optimised Federated Learning Efficiency}: We enhance the FL process by minimising communication overhead, improving scalability, and maintaining privacy-preserving model aggregation.
\end{itemize}

Therefore, our introduced framework (FedORA) integrates FL with ORAN's adaptive capabilities, enabling efficient load balancing across different RATs. The ORAN-based rApp/xApp framework continuously monitors network conditions, analyses traffic patterns, and optimises resource allocation to sustain high FL performance with reduced energy costs.

\begin{figure}[t]
    \centering
    \includegraphics[width=.8\linewidth, height=0.250\textheight]{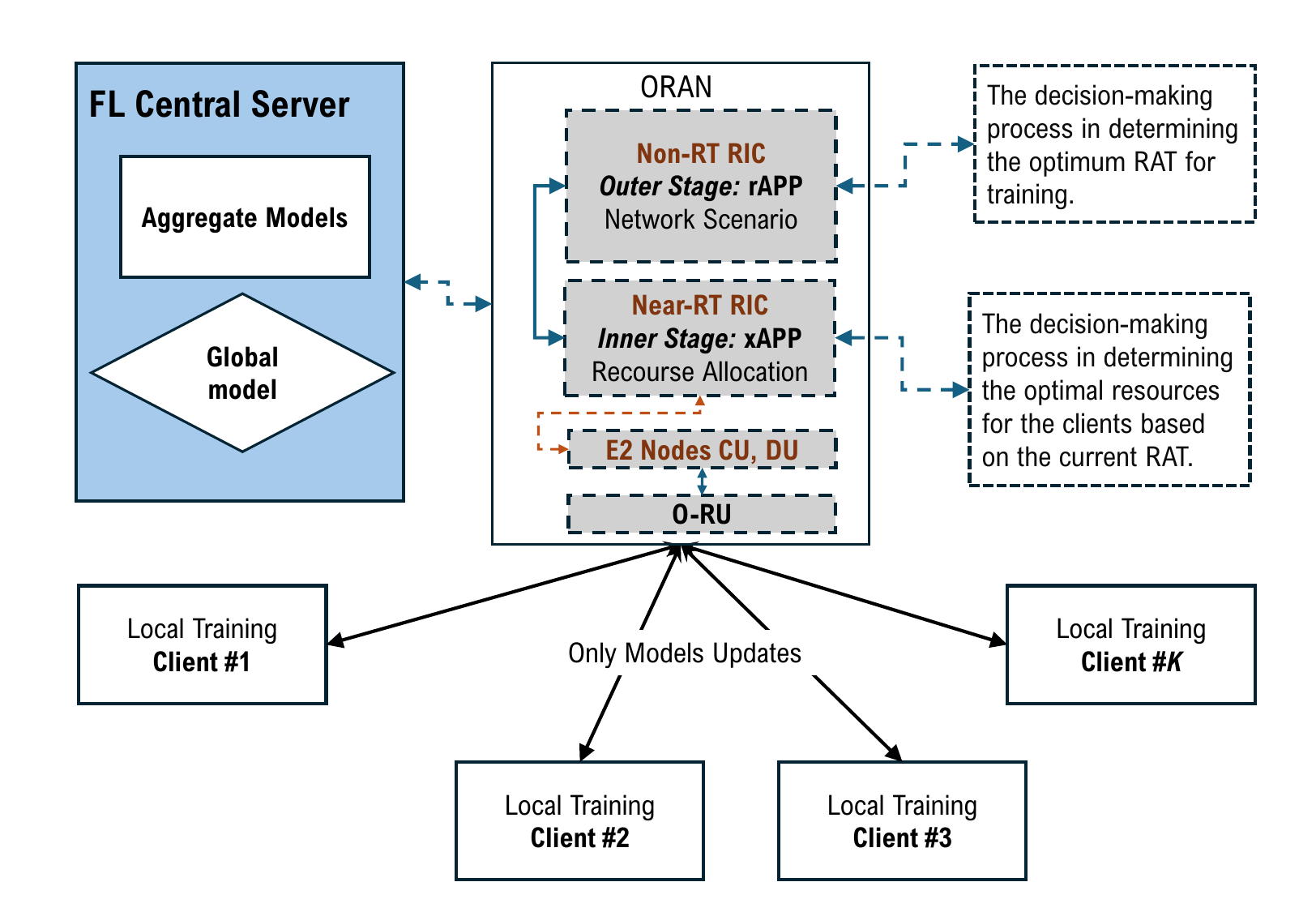}
    \caption{The proposed FedORA Network layout}
    \label{fig:ORAN FL}
\end{figure}

\subsection{ORAN and RATs-Enabled Network Configuration}
In our proposed model, each user device is represented as a distributed node within an ORAN-based network, configured with RIC to facilitate data transmission over various RATs. ORAN's modular structure allows it to disaggregate traditional gNB components into three primary units: the Open Radio Unit (O-RU), the Open Distributed Unit (O-DU), and the Open Centralised Unit (O-CU) \cite{qazzaz2025oran}. These units are strategically deployed across both the base station and edge cloud, enabling efficient model parameter transmission and flexible device-server communication within FL environments.

\subsection{Two-Stage Optimisation Strategy Using rApp and xApp}
\label{sub:Two-stage}

To optimise the network resources usage in ORAN networks for the FL framework, we present a two-stage optimisation approach that leverages ORAN's rApp and xApp functionalities:

\begin{itemize}
   
    \item \textbf{Stage 1 (Outer Stage)}: The rApp model in the Non-RT RIC employs Reinforcement Learning (RL) to optimise RAT selection and power allocation for FL users by monitoring network KPIs such as latency and QoS.

    \item \textbf{Stage 2 (Inner Stage)}: In this phase, xApp algorithms deployed on the Near-Real-Time RIC facilitate near-real-time optimisation of pathway selection, enabling the most efficient data transmission routes for FL model parameters. This stage employs model-based wireless networks for resource allocation. 

\end{itemize}

\section{Simulation Results and Discussion} \label{sec: Simulation and results}

We benchmarked our learning model against related and well-known FL algorithms, including Google's FedAvg model \cite{mcmahan2017communicationefficient} and other state-of-the-art approaches such as FLAIR \cite{sharma2023flair} and Greedy FL \cite{mehta2023greedy}. The FL learning system leverages Python and Pytorch APIs to train users to optimise a classification problem on the widely recognised CIFAR-10 dataset comprising 60,000 colour images categorically distributed across 10 distinct image classifications.

\begin{figure}[!t]
    \centering
    \includegraphics[width=1.03\linewidth, height=0.25\textheight]{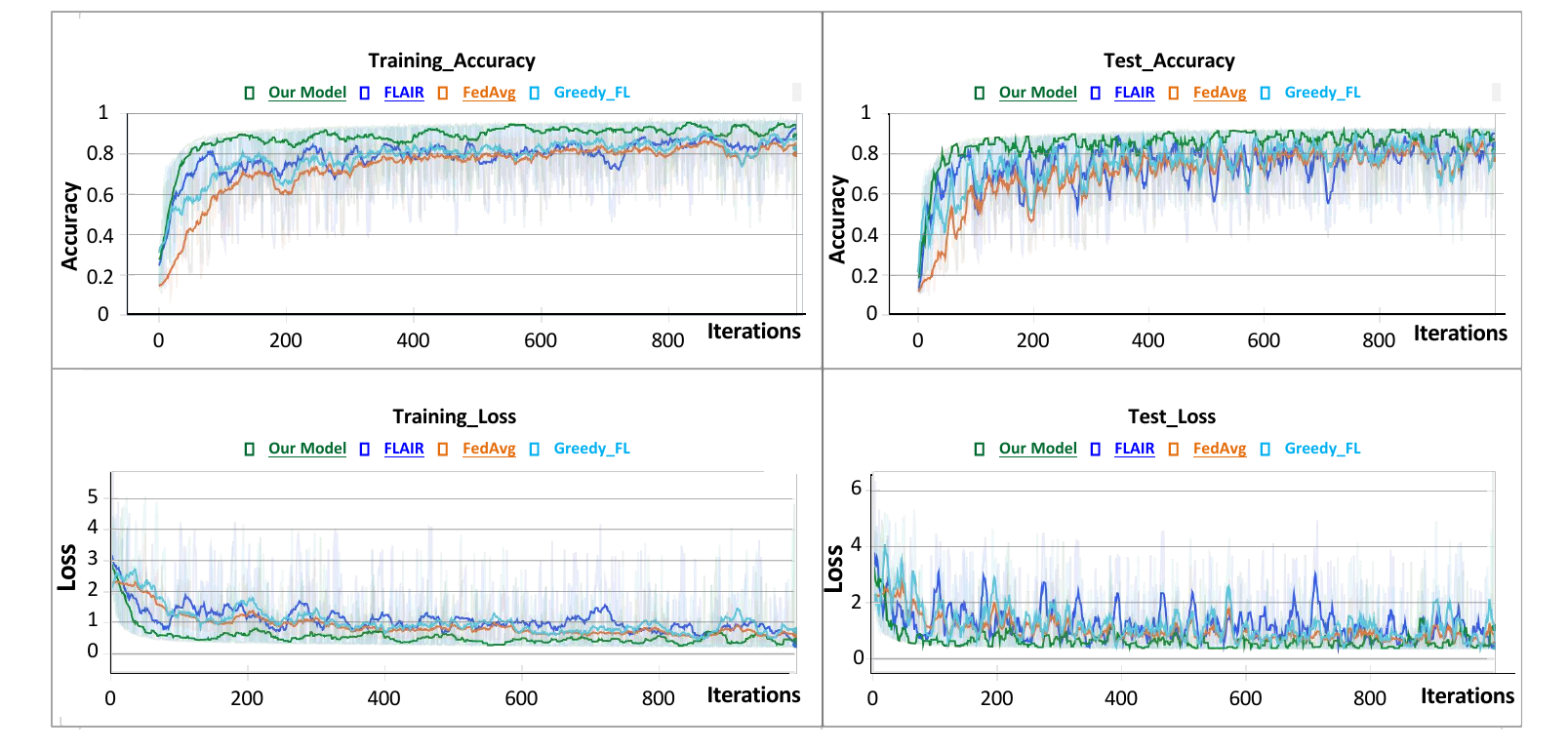}
    \caption{Comparative performance of our model's accuracy and loss against other relevant models.}
    \label{fig:MPTCP_accuracy}
\end{figure}

\begin{table}[!t]
    \centering
    \scriptsize 
    \setlength{\tabcolsep}{3.5pt} 
    \renewcommand{\arraystretch}{1.1} 
    \begin{tabular}{cccccccc} 
        \toprule
        \textbf{Model} & \textbf{\shortstack{Client\\Alg.}} & \textbf{Dataset} & \textbf{Acc.} & \textbf{Loss} & \textbf{\shortstack{Std.\\Dev.}} & \textbf{\shortstack{FL Net.\\QoS}} & 
        \textbf{\shortstack{Avg. Power\\(Watt)}} \\
        \midrule
        Our Model & CNN & CIFAR-10 & \textbf{96.5} & \textbf{0.038} & \textbf{0.15} & \checkmark & \textbf{20.08} \\
        \hline
        FLAIR & CNN & CIFAR-10 & 94.1 & 0.042 & 0.32 & \sffamily x & 24.80  \\
        \hline
        FedAvg & CNN & CIFAR-10 & 85.3 & 0.085 & 0.40 & \sffamily x & 24.80  \\
        \hline
        Greedy FL & CNN & CIFAR-10 & 90.2 & 0.093 & 0.29 & \sffamily x & 24.80  \\
        \bottomrule
    \end{tabular}
    \caption{Comparison of our FedORA model using ORAN architecture against other models' performance and system metrics.}
    \label{tab:DFL_MPTCP}
\end{table}

Our framework (FedORA), enhanced by the proposed resource allocation model, improves the real-world applications' requirements, such as reliability and scalability, that lead to achieving robust and competitive model accuracy compared to traditional FL models, as shown in Figure \ref{fig:MPTCP_accuracy}. The energy consumption remains low as our model dynamically selects the optimal resource allocation policy, ensuring that both outage constraints and the QoS requirements for the FL model are satisfied. To this end, the efficient management of resources within the proposed FL network creates a robust and comprehensive framework that enables clients to use the available resources effectively. Consequently, our approach is well-suited for scalable, collaborative learning applications that require sensitivity to energy efficiency and network performance.

\section{Conclusion} \label{sec: conclusion}

In this work, our FedORA framework achieves highly competitive results with low power consumption compared to existing advanced models, demonstrating its suitability for real-time applications requiring accuracy and efficiency. This robust and comprehensive framework empowers clients to effectively leverage available resources, making it ideal for scalable, collaborative learning applications that prioritise privacy, energy efficiency and network performance.

\section*{Acknowledgement}
This research was funded by EP/X040518/1 EPSRC CHEDDAR and was partly funded by UKRI Grant EP/X039161/1 and MSCA Horizon EU Grant 101086218.

\bibliographystyle{IEEEtran}
\bibliography{references}

\end{document}